\documentclass[traditabstract]{aa}

\usepackage{amsmath}
\usepackage{amssymb}
\usepackage{graphicx}
\usepackage{txfonts}
\usepackage{dcolumn}
\usepackage{natbib}
\usepackage{url}

\bibpunct{(}{)}{;}{a}{}{,}

\newcommand{\diff}{{\textrm{d}}}
\newcommand{\Diff}{{\textrm{D}}}
\newcommand{\trm}[1]{{\textrm{#1}}}

\begin{document}
 \title{Overshooting by convective settling}

 \author{R. Andr\'assy, H. C. Spruit}

 \institute{Max Planck Institute for Astrophysics, Karl-Schwarzschildstr. 1, 85748 Garching, Germany}

 \date{Received ; accepted }

 \abstract{
We study a process of slow mixing in stars with convective envelopes, which is driven by the settling of cool downward plumes below the base of the convection zone. If a small fraction (of order $10^{-7}$) of the material cooled at the surface retains a significant entropy deficit while descending in plumes, it can reach the depth where lithium burning takes place. The model calculates the thermal response and mixing below the convection zone due to the settling process, assuming that the plumes arrive at the base of the convection zone with a broad range of entropy contrasts. We obtain a good fit to the observed lithium depletion in the Sun by assuming that the settling mass flux is distributed with respect to the entropy contrast as a power law with a slope around -2. We find convective settling to have a negligible influence on the stratification below the convection zone, although mixing induced by it could modify the gradient of helium concentration.
 }

 \maketitle
%
%________________________________________________________________

% **********************************************************************
% *                         Introduction                               *
% **********************************************************************
\section{Introduction}
\label{sec:introduction}

In stellar evolution, the term `overshooting' is used to denote any process that can extend the mixing effect of a convective flow beyond the region of linear instability of the stratification. A number of conceptually different processes have been considered, operating on a range of length scales and time scales.

The first overshooting mechanism one can think of is the extension of the convection zone as caused by the inertia and entropy contrast of the convective flows, which they developed before leaving the convection zone proper. \citet{Roxburgh65} and \citet{SaslawSchwarzschild65} both estimated the expected extent of overshooting due to this process using characteristic values of the speed and entropy contrast predicted by the mixing-length theory (MLT). They obtained a negligible penetration distance as a consequence of the steep entropy gradient in the radiative zone and low convective velocities connected to the efficiency of deep convection in stars.

This process can be called `overshooting' in the literal (ballistic) sense, operating on a short time scale and over a negligible distance. It was recognised early on \citep[e.g.][]{ShavivSalpeter73} that the convective cells at the boundary with a stable zone could also have a slower but systematic effect that extends the region of convectively overturning motion. Such effects would be much more relevant on the longer time scales of stellar evolution. To distinguish it from the ballistic process, this kind of process can be called `convective penetration' \citep{Zahn91}. Models have been developed by several authors. \Citet{vanBallegooijen82} studied the long-term response of the sub-adiabatic interior to a stationary convective flow at the base of the convection zone. The model predicted a mildly sub-adiabatic overshoot region terminated by a thin boundary layer. Related ideas were developed by \citet{SchmittEtal84} and \citet{Rempel04} using models for the interaction of downward plumes with their environment.

For the discussion of overshooting/penetration it is useful to make a distinction between the case of a convective envelope like the Sun and Sun-like stars and the conditions in internal convective zones in stars. Owing to the very low gas density at the stellar surface, where the flows are driven, the flows in a convective envelope are far more `plume-like' than in a convective core, where the density stratification is much less extreme. This has major consequences for the overshooting problem. Whereas in core convection the order of magnitude of velocities and temperature fluctuations can be plausibly estimated from a mixing-length formula, the downward plumes in a convective envelope have much stronger entropy contrasts than mixing-length estimates based on some average of the stratification. To the (as yet poorly known) extent to which these plumes survive mixing, entrainment and merging (see e.g. \citealp{Viallet12}), they will arrive at the base of the convection zone with a broad range of entropy contrasts. This mixture will settle in the stable layers below at a range of depths, which are determined by the distribution of entropy contrasts (see \citealp{NordlundStein95} for an example of a simulation showing the effect qualitatively). We call this form of gentle overshooting `convective settling' (shortened to `settling' hereinafter). The explicit inclusion of the cold plumes makes it a rather different contribution to overshooting compared with the processes mentioned above. On the long time scales relevant to stellar evolution, it has the potential of producing a weak mixing extending to deeper layers.

Numerical simulations are naturally restricted to limited time scales that cover the ballistic process much better than the slower processes of penetration and convective settling. Extrapolations have to be made to translate simulation results from a numerically accessible regime to astrophysically relevant conditions. Since different contributing processes act on different time scales and depths, such extrapolation would require disentangling them from a numerical simulation. This is not a straightforward task. Early two-dimensional computations by \citet{HurlburtEtal86} showed strong downward-directed plumes penetrating deep into the stable stratification and generating gravity waves there. Later studies \citep[e.g.][]{HurlburtEtal94, SinghEtal95, BrummellEtal02, RogersGlatzmaier05} focussed on the dependence of penetration depth on the stiffness of the interface separating the stable and the unstable stratification. With 3D simulations of a convective envelope model, \citet{NordlundStein95} showed qualitatively how the transition from convection to the stable interior is softened by the spread in entropy of the downward plumes, i.e. the `settling' process above. Realistic simulations of a convective envelope are still quite outside the accessible domain. This is due to the daunting range of length and time scales needed to cover the driving at the stellar surface, the mixing and entrainment in the plumes, and their compression over the enormous density range between the surface and the base \citep[cf.][]{RogersGlatzmaier05, RogersEtal06}.

A further phenomenon that could lead to mixing at the bottom of such a convection zone is a meridional circulation in the radiative zone, driven by the latitudinal differential rotation of the convective envelope. This was studied by \citet{SpiegelZahn92}, among others, who calculated how viscous stress causes the pattern of differential rotation to penetrate the stable stratification on a long time scale.

\citet{McIntyre07} points out that the problem is identical to the so-called `gyroscopic pumping' process \citep{HaynesEtal91}, in which thermal diffusion plays a dominant role. \citet{GaraudArreguin09} show how this process is capable of producing meridional flows penetrating deep into the radiative interior, provided that there is a source of stress in it.

Stochastic convective motions generate a whole spectrum of internal gravity waves propagating into the stable stratification. These can in principle also produce a modest amount of mixing at some distance away from a convection zone, see e.g. \citet{Press81, GarciaSpruit91, Schatzman96}.

An important clue about the deeper, slow mixing processes is the long-term lithium depletion observed in low-mass main-sequence stars \citep[][and references therein]{Herbig65, Boesgaard76, Pinsonneault94}. From the complex overshooting physics described above, we decided to isolate the settling process since it could be the most relevant for lithium depletion as suggested before in \citet{Spruit97}. We present the physics of settling and a simplified model of it in Sec.~\ref{sec:model_physics}, then we formulate the model mathematically in Sec.~\ref{sec:mathematical_formulation} and describe our approach to estimating the extent of lithium depletion in Sec.~\ref{sec:lithium_burning}. We focus on the solar case, because we can also apply helioseismic constraints there \citep{DalsgaardEtal96, BahcallEtal98, SchlattlWeiss99}. Results of our study are summarised in Sec.~\ref{sec:results} and discussed in Sec.~\ref{sec:discussion}.

% **********************************************************************
% *                              Model                                 *
% **********************************************************************
\section{Model}
\label{sec:model}

As discussed above, several processes are present at the same time at the boundary of a stellar convection zone and act on a range of depths and time scales. We focus on lithium depletion here, so we need to consider mixing processes working deep in the stable stratification ($T \gtrsim 2.5$~MK) on nuclear-burning time scales. We envisage a mechanism based on the idea that, apart from the MLT-like flows, there might be a tiny fraction of substantially colder material at the lower boundary of an envelope convection zone. It would be a (partially mixed) remnant of the photospheric downflows. Its low entropy would make it sink until it has reached neutral buoyancy. Because of the large entropy deficit, with which the plumes start, this can in principle happen rather deep in the stable stratification. One would expect, however, that the larger the entropy deficit, the smaller the fraction of the downflowing material involved. We expect the settling rate at lithium-burning depths to be so low that settling negligibly affects the stratification there, but still high enough to ensure significant mass exchange with the convection zone on very long time scales. On the other hand, the settling rate in the vicinity of the convection zone could be much greater, possibly leaving a footprint on the sound-speed profile.

Settling itself involves a vast range of time scales. The downflows sink towards the settling point on their dynamical time scale. Since only a small fraction of the plumes will survive with a significant entropy deficit, the filling factor of this fraction is small. This implies that the upflow enforced by mass conservation is very slow. This circulation disturbs the radiative equilibrium, which in a steady state is balanced by radiative diffusion. It is a certain time average of this equilibration process that represents the influence of settling on the stratification on the time scale of the star's lifetime.

What fraction of the photospheric downflows should keep their low entropy? To get an order-of-magnitude estimate from the observed value of the Sun's lithium depletion, assume that a fraction $\epsilon$ of the photospheric downward mass flux $\mathcal{F}_\trm{ph}$ sinks deep enough to reach lithium-burning conditions, burning its lithium content instantaneously there. To burn all the observable lithium, the whole mass of the convection zone $M_\trm{cz}$ (per unit area) must be replaced by the flux $\epsilon\,\mathcal{F}_\trm{ph}$. If this is to happen on the time scale $\tau$, $\epsilon = M_\trm{cz}/(\mathcal{F}_\trm{ph}\tau)$. This way we obtain $\epsilon \approx 10^{-7}$ for solar values ($M_\trm{cz} \approx 7\times 10^8$\,g\,cm$^{-2}$, $\mathcal{F}_\trm{ph} \approx 3\times 10^{-2}$\,g\,cm$^{-2}$\,s$^{-1}$ and $\tau \approx 5\times 10^9$\,yr). Such a minute amount would hardly be traceable by current ab initio simulations.

% **********************************************************************
% *                  Model physics and simplifications                 *
% **********************************************************************
\subsection{Model physics and simplifications}
\label{sec:model_physics}

We have constructed a simplified one-dimensional model of settling (Sec.~\ref{sec:mathematical_formulation}) based on the ideas described above. It includes three assumptions. First, we ignore any exchange of heat or mass between the downflows and the upflow within the region bounded by the Schwarzschild boundary at the top and the point where the given downflow settles at the bottom (i.e. all downflows are adiabatic in our model). Secondly, we do not consider the kinetic energy of the downflows and its dissipation. Third, we regard the overshoot region as chemically homogeneous (with the exception of lithium). The last assumption has a significant influence on the sound-speed profile and is justified a posteriori in Sec.~\ref{sec:results}.

Since we are only interested in very long time scales, the model is formulated as stationary --- we base it on the equality of a local cooling rate due to advection (i.e. settling) and a local heating rate due to a negative radiative flux divergence. The very nature of settling, in which cold material sinks downwards and pushes hot material upwards, causes the convective flux to be positive, which is unusual for overshooting models. Conservation of the total flux then makes the radiative flux decrease, which is achieved by a decrease in the temperature gradient $\nabla$. This immediately tells us that a model with settling will be warmer than the one without it. The decrease in $\nabla$ will also shift the formal boundary of convective instability in the settling model higher up into the stratification and slightly {\em reduce} the depth of the convection zone. The settling process adds mass from the convection zone to the stratification below, which responds by a slow upward flow, eventually returning mass to the convection zone. The location where the mass settles depends on its entropy since it is the depth at which it matches that of the stratification. The model calculates how settling changes the steady state stratification under the combined effects of thermal diffusion and the assumed settling mass flux. 

The model thus needs a description for the flux of settling mass as a function of its entropy. Since current numerical simulations still cannot reach the parameter values needed to get this distribution from first principles, it needs to be parametrised. We describe it with a power law. We define the mass flux (either upflow or downflow) in terms of the quantity $\eta\rho v$, where $\eta$ is the filling factor (the relative geometrical area covered by the flow, $0 \le \eta \le 1$), $\rho$ is the density, and $v$ the velocity. We avoid any explicit usage of the unknown values of $v$ and $\eta$ this way. (The model only requires $\eta \ll 1$ for the downflow, see above.) The range of entropies in the mass flux distribution is bounded between the entropy of a typical downflow in the photosphere and the MLT estimate of the entropy contrast in the lower part of the convection zone.

% **********************************************************************
% *                     Mathematical formulation                       *
% **********************************************************************
\subsection{ Mathematical formulation}
\label{sec:mathematical_formulation}

To describe the settling process mathematically, we model the settling region as a plane parallel layer (instead of a spherical shell) of ideal gas in a homogeneous gravitational field. These assumptions are not essential but make the mathematics more transparent and are not likely to change the outcomes substantially.

We describe the thermodynamic state by dimensionless variables $\hat{p}$, $\hat{\rho}$, $\hat{T}$, and $\hat{s}$, which stand for the pressure, density, temperature, and specific entropy, respectively:
\begin{align}
\hat{p} &= \frac{p}{p_0}, \\
\hat{\rho} &= \frac{\rho}{\rho_0}, \\
\hat{T} &= \frac{T}{T_0}, \label{eq:T_hat} \\
\hat{s} &= \frac{s - s_0}{\mathbb{R}}, \label{eq:s_hat}
\end{align}
where $p$, $\rho$, $T$, and $s$ are the physical state variables, the index zero corresponds to a reference point, and $\mathbb{R}$ is the gas constant. We use $\mathbb{R} \equiv k_\trm{B}/(\mu\,m_\trm{p})$, so that it includes the mean molecular weight $\mu = \textrm{const.}$ (see Sec.~\ref{sec:model_physics}). The reference point, which must not be influenced by settling, can be put anywhere beneath the settling layer. The thermodynamic state we refer to is a horizontal average set by the upflow state, because the downflow filling factor is assumed to be small. We describe the thermodynamics of settling in terms of this average state.

We introduce a dimensionless independent variable
\begin{equation}
\psi = -\ln(p/p_0),
\label{eq:psi_definition}
\end{equation}
which can be related to the geometrical height $z$ by using the condition of hydrostatic equilibrium,
\begin{equation}
dp = -\rho\,g\,dz,
\label{eq:hydrostatic_equilibrium}
\end{equation}
where $g = \trm{const.}$ is the gravitational acceleration, and we set $z = 0$ at the reference point. If we differentiate Eq.~\ref{eq:psi_definition}, combine it with Eq.~\ref{eq:hydrostatic_equilibrium}, and use the ideal gas law $p = \rho\,\mathbb{R}\,T$, we obtain a recipe for converting derivatives to the dimensionless form, 
\begin{equation}
\frac{\diff}{\diff \psi} = \frac{\mathbb{R}T}{g} \frac{\diff}{\diff z}.
\label{eq:gradient_transformation}
\end{equation}
The local pressure scale height is $H=\mathbb{R}T/g$. Elementary thermodynamics provide us with the following useful relations 
\begin{align}
\hat{\rho} &= \hat{p}^{1 - \nabla_\trm{ad}} \exp\ (-\nabla_\trm{ad} \hat{s}), \label{eq:rho_hat_vs_s_hat} \\
\hat{T} &= \hat{p}^{\nabla_\trm{ad}} \exp\ (\nabla_\trm{ad} \hat{s}), \label{eq:t_hat_vs_s_hat}\\
\frac{\diff \hat{s}}{\diff \psi} &= 1 - \frac{\nabla}{\nabla_\trm{ad}}, \label{eq:nabla}
\end{align}
where $\nabla_\trm{ad}$ is the usual adiabatic temperature gradient. 

Let $\diff\dot m(\hat s)$ be the differential settling mass flux per unit of entropy $\hat s$ at the base of the convection zone. Its functional form is assumed to be a power law,
\begin{align}
\diff\dot{m} &= -\dot{M} f(\hat{s})\,\diff\hat{s}, \label{eq:mass_flux_distribution}\\
f(\hat{s}) &= N\,\left(\frac{\hat{s}_\trm{sb} - \hat{s}}{(\delta \hat{s})_\trm{min}}\right)^{-\beta}, \label{eq:distribution_function}
\end{align}
where $\dot{M} > 0$ is the total settling mass flux, $f(\hat{s})$ is a distribution function, $N$ a normalisation factor, $\hat{s}_\trm{sb}$ the entropy of the stratification at the Schwarzschild boundary (where $\nabla = \nabla_\trm{ad}$ and $\diff\hat{s}/\diff\psi = 0$, see Eq.~\ref{eq:nabla}), $(\delta \hat{s})_\trm{min} > 0$ is the lowest entropy contrast of a downflow with respect to $\hat{s}_\trm{sb}$, and $\beta > 0$ describes the steepness of the distribution. We also define a maximum entropy contrast $(\delta \hat{s})_\trm{max}$, which is determined by the entropy of the downflows at the surface (see Sec.~\ref{sec:model_physics}), and set $f(\hat{s}) = 0$ for $\hat{s} < \hat{s}_\trm{sb} - (\delta \hat{s})_\trm{max}$ and $\hat{s} > \hat{s}_\trm{sb} - (\delta \hat{s})_\trm{min}$. We require that the area under $f(\hat{s})$ be unity, so that (for $\beta \neq 1$) 
\begin{equation}
N = \frac{\beta - 1}{(\delta \hat{s})_\trm{min}} \left[1 - \left(\frac{(\delta \hat{s})_\trm{max}}{(\delta \hat{s})_\trm{min}}\right)^{-(\beta - 1)}\right]^{-1}.
\end{equation}
We define a cumulative distribution function
\begin{equation}
F(\hat{s}) = 1 - \int\limits_{\hat{s}}^\infty f(\hat{s}^\prime)\ \diff\hat{s}^\prime,
\label{eq:cumulative_distribution_function}
\end{equation}
which describes the relative amount of downflows that settle \emph{below} the level where the entropy in the stratification equals $\hat{s}$. All downflows of entropy $\hat{s}^\prime > \hat{s}$ have already reached neutral buoyancy and settled higher up ($\diff\hat{s}/\diff{z} > 0$ for $\nabla < \nabla_\trm{ad}$, see Eqs.~\ref{eq:gradient_transformation} and \ref{eq:nabla}). Therefore \mbox{$F(\hat{s}) = 0$} for \mbox{$\hat{s} \le \hat{s}_\trm{sb} - (\delta \hat{s})_\trm{max}$}, \mbox{$0 < F(\hat{s}) < 1$} for \mbox{$\hat{s}_\trm{sb} - (\delta \hat{s})_\trm{max} < \hat{s} < \hat{s}_\trm{sb} - (\delta \hat{s})_\trm{min}$}, and \mbox{$F(\hat{s}) = 1$} for \mbox{$\hat{s} \ge \hat{s}_\trm{sb} - (\delta \hat{s})_\trm{min}$}.

We write the energy equation in the upflow in terms of entropy, in the Lagrangian form
\begin{equation}
\rho\,T \frac{\Diff s}{\Diff t} = -\frac{\diff \mathcal{F}_\trm{rad}}{\diff z},
\label{eq:energy_equation}
\end{equation}
where $t$ is the time, $\mathcal{F}_\trm{rad}$ the radiative flux, \mbox{$\Diff/\Diff t = \partial/\partial t + v\:\diff/\diff z$} is the Lagrangian time derivative, and $v$ the upflow velocity.

Global mass conservation requires the total amount of mass being transported downward to be equal to the total amount of mass being transported upward through any surface $z = \trm{const.}$ With our approximation that the filling factor of the upflow is close to unity, the upward mass flux $\dot{M}\,F(\hat{s})$ is given by the product of the upflow density and velocity,
\begin{equation}
\dot{M}\,F(\hat{s}) = \rho\,v.
\label{eq:upflow_velocity}
\end{equation} 
This allows us to formulate the model without the knowledge of filling factors or using a momentum equation.

The stationary nature of our model eliminates the $\partial/\partial t$ term in Eq.~\ref{eq:energy_equation}, and using Eq.~\ref{eq:upflow_velocity}, we can write the upflow energy balance in the form
\begin{equation}
\dot{M}\,F(\hat{s})\,T \frac{\diff s}{\diff z} = -\frac{\diff \mathcal{F}_\trm{rad}}{\diff z}.
\label{eq:energy_equation2}
\end{equation}
One can obtain a dimensionless form of this equation by using Eqs.~\ref{eq:T_hat}, \ref{eq:s_hat}, and \ref{eq:gradient_transformation} and introducing a dimensionless radiative flux $\hat{\mathcal{F}}_\trm{rad} = \mathcal{F}_\trm{rad}/\mathcal{F}_\trm{tot}$, where $\mathcal{F}_\trm{tot}$ is the total flux (being equal to the radiative one at the reference point). The state variables can be related to the entropy by Eqs.~\ref{eq:rho_hat_vs_s_hat} and \ref{eq:t_hat_vs_s_hat}. The diffusive approximation of the radiative flux is
\begin{equation}
\mathcal{F}_\trm{rad} = \frac{16}{3} \frac{g\,\sigma T^4}{p\,\kappa} \nabla.\label{eq:frad}
\end{equation}
With Eq.~\ref{eq:nabla}, Eq.~\ref{eq:energy_equation2} then yields
\begin{equation}
C\,\hat{T}(\psi, \hat{s}) \frac{\diff \hat{s}}{\diff \psi}F(\hat{s}) = -\frac{\diff \hat{\mathcal{F}}_\trm{rad}(\psi, \hat{s}, \diff\hat{s}/\diff{\psi})}{\diff \psi},
\label{eq:settling_equation}
\end{equation}
where 
\begin{equation}
C = \frac{\dot{M}\,\mathbb{R}T_0}{\mathcal{F}_\trm{tot}}
\end{equation}
is a ratio of a characteristic convective flux to the star's net energy flux ${\mathcal{F}_\trm{tot}}$ at the base of the convection zone. It is a dimensionless measure of the settling mass flux $\dot{M}$. The explicit form of the derivative $\diff \hat{\mathcal{F}}_\trm{rad}/\diff \psi$ on the right-hand side of Eq.~\ref{eq:settling_equation} is
\begin{align}
\frac{\diff \hat{\mathcal{F}}_\trm{rad}}{\diff \psi} = &\frac{\nabla_\trm{ad}}{\nabla_0}\hat{\kappa}^{-1}\exp\left[(1 - 4\nabla_\trm{ad})\psi + 4\nabla_\trm{ad}\hat{s}\right]\times\nonumber \\
&\Bigg[1 - 4\nabla_\trm{ad} - \frac{\partial \ln\hat{\kappa}}{\partial \psi} + \left(8\nabla_\trm{ad} - 1 + \frac{\partial \ln\hat{\kappa}}{\partial \psi} - \frac{\partial \ln\hat{\kappa}}{\partial \hat{s}}\right)\frac{\diff\hat{s}}{\diff\psi} - \nonumber \\
&\left(4\nabla_\trm{ad} - \frac{\partial \ln\hat{\kappa}}{\partial \hat{s}}\right)\left(\frac{\diff\hat{s}}{\diff\psi}\right)^2 - \frac{\diff^2\hat{s}}{\diff\psi^2}\Bigg],
\label{eq:dfraddpsi}
\end{align}
where $\hat{\kappa} = \kappa/\kappa_0$ is the dimensionless opacity function and $\kappa_0$ the opacity value at the reference point.

Equation \ref{eq:settling_equation} governs the whole settling process in our model. It is a non-linear, second-order ordinary differential equation for the entropy profile $\hat{s}(\psi)$. The opacity function $\hat{\kappa}(\psi,\,\hat{s})$ in Eq.~\ref{eq:dfraddpsi} could by specified by standard opacity tables, but that is not necessary in such a simplified model. In our sample calculations presented in Sec.~\ref{sec:results}, we used the opacity law
\begin{equation}
\ln\hat{\kappa} = \alpha\,\psi,
\end{equation}
where the exponent $\alpha$ is a fitting parameter. In this simplification the opacity depends only on the pressure (via $\psi$) and not on the full thermodynamic state. This prescription is sufficient for our purposes if it reasonably fits the opacity profile of the stratification without settling, and if the state change due to settling is small.

Equation~\ref{eq:settling_equation} implicitly contains an a priori unknown value of the entropy at the Schwarzschild boundary, $\hat{s}_\trm{sb}$, as an input parameter of the distribution function $F(\hat{s})$, see Eqs.~\ref{eq:distribution_function} and \ref{eq:cumulative_distribution_function}. Therefore any solution procedure must involve iterations. One could pick an initial guess $\hat{s}_\trm{sb}^{\,0}$ and integrate Eq.~\ref{eq:settling_equation} from the reference point (where $\psi = 0$, see Eq.~\ref{eq:psi_definition}) upwards. The initial conditions would be $\hat{s} = 0$, $\diff\hat{s}/\diff\psi = 1 - \nabla_0/\nabla_\trm{ad}$ (see Eqs.~\ref{eq:s_hat} and \ref{eq:nabla}), where $\nabla_0$ is the known temperature gradient at the reference point. The integration would then be stopped at the point $\psi_\trm{sb}^{\,0}$, where the entropy reaches a maximum (i.e. the Schwarzschild boundary). The solution value at this point, $\hat{s}^{\,0}\left(\psi_\trm{sb}^{\,0}\right) \neq \hat{s}_\trm{sb}^{\,0}$ (in general), could be used as a new estimate $\hat{s}_\trm{sb}^{\,1}$, and the whole process could be repeated until convergence. In reality, the steep profile of $F(\hat{s})$ renders this method highly unstable.

Our sample calculations shown in Sec.~\ref{sec:results} were computed by modifying this method. We started the \mbox{$i$-th} iteration by integrating Eq.~\ref{eq:settling_equation} from an estimated position of the Schwarzschild boundary $\left(\psi_\trm{sb}^{\,i}\,,\ s_\trm{sb}^{\,i}\right)$ downwards to the reference point. The initial conditions were $\hat{s} = s_\trm{sb}^{\,i}$, $\diff\hat{s}/\diff\psi = 0$ (see Eq.~\ref{eq:nabla}). We stopped the integration at $\psi_0^{\,i}$ such that $\hat{s}^{\,i}\left(\psi_0^{\,i}\right) = 0$. This, in general, leads to \mbox{$\psi_0^{\,i} \neq 0$} and \mbox{$\diff\hat{s}/\diff\psi\big|\left(\psi=\psi_0^{\,i}\right) \neq 1 - \nabla_0/\nabla_\trm{ad}$}; i.e., the solution curve misses the reference point. Therefore we make a new estimate $\left(\psi_\trm{sb}^{\,i+1}\,,\ s_\trm{sb}^{\,i+1}\right)$ and repeat the procedure until the solution passes close enough to the reference point, and the entropy gradient gets close enough to $1-\nabla_0/\nabla_\trm{ad}$ there. This method converges smoothly to the desired solution.

% **********************************************************************
% *                         Lithium burning                            *
% **********************************************************************
\subsection{Lithium burning}
\label{sec:lithium_burning}

The extent of lithium depletion in the convection zone is given by two time scales, which are both strong functions of depth. Taking a horizontal layer of thickness $\diff z$, we introduce a `recycling' time scale $\tau_\trm{r}$ by defining the mass exchange rate (per unit area) in this layer as $\rho\,\diff z/\tau_\trm{r}$. The height interval $\diff z$ corresponds to an entropy interval $\diff \hat{s}$ in the stratification, which implies that the mass exchange rate due to settling is $\dot{M} f(\hat{s}) \diff\hat{s}$ (see Eq.~\ref{eq:mass_flux_distribution}). Equating the last two expressions, we obtain
\begin{equation}
1/\tau_\trm{r} =\frac{\dot{M}}{\rho}f(\hat{s})\, \frac{\diff \hat{s}}{\diff z}=
\frac{g}{p}\, \dot{M}f(\hat{s})\, \frac{\diff \hat{s}}{\diff \psi},
\label{eq:tau_r}
\end{equation}
where we have also used Eq.~\ref{eq:gradient_transformation} and the equation of state. 

The second time scale describes the speed of the burning itself. Lithium is burned by the reaction $^7$Li(p, $\alpha$)$\alpha$, which causes a decrease in its abundance $A \equiv N_{\rm Li}/N_\trm{H}$ on the time scale \mbox{$\tau_\trm{b} \equiv -\left(\diff \ln A / \diff t\right)^{-1}$}. With the reaction's astrophysical S-factor \mbox{$S_\trm{b}(0) = 55$\,keV\,barns} from \citet{LattuadaEtal01}, the burning time scale is \citep[see e.g.][]{HansenKawaler94}
\begin{equation}
\tau_\trm{b} = \left[9.02\times 10^6\, X\rho\ \xi^2\exp(-\xi)\right]^{-1}\ \textrm{yr},
\label{eq:tau_b}
\end{equation}
where $X$ is the hydrogen mass fraction, the density $\rho$ is in g\,cm$^{-3}$, and $\xi = 84.5\, T_6^{-1/3}$ with the temperature $T_6$ in MK. 

We discretise the settling layer into a grid of $n$ sub-layers ordered by height. The bottom of the $i$-th sub-layer is located at a height of $z_i$, $i=1,2,\dots,n$. The resolution of the grid is chosen by setting a maximum to the relative changes in $F[\hat{s}(z)]$, $\tau_\trm{r}(z)$ and $\tau_\trm{b}(z)$ between grid points. We put the first grid point to the maximal depth settling can reach, i.e. $z_1 = \max\{z\colon F[\hat{s}(z)]=0\}$. The topmost point represents the convection zone itself including a well-mixed upper part of the overshoot region, where the burning rate is negligible. Lithium is assumed to be a trace element.

We model the burning process by the set of equations
\begin{equation}
\frac{\diff N_i}{\diff t} = R_{\trm{b},i} + R_{\trm{s},i} + R_{\trm{a},i},
\label{eq:burning_equations}
\end{equation}
where $N_i$ is the number density of lithium atoms per unit area in the $i$-th sub-layer, $t$ is the time, $R_{\trm{b},i}$ a burning rate, $R_{\trm{s},i}$ a settling rate of `fresh' lithium atoms from the convection zone, and $R_{\trm{a},i}$ a rate of lithium transport by advection. Nuclear burning is an exponential decay process, hence
\begin{equation}
R_{\trm{b},i} = -\frac{N_i}{\tau_{\trm{b},i}},
\label{eq:rbi}
\end{equation}
where $\tau_{\trm{b},i}$ can be defined e.g. as \mbox{$\tau_{\trm{b},i} = [\tau_\trm{b}(z_i) + \tau_\trm{b}(z_{i+1})]/2$} with a special case $\tau_{\trm{b},n} = 0$ (see above). We introduce the mass settling rate $\dot{m}_i$ in the $i$-th sub-layer (cf. Eq.~\ref{eq:cumulative_distribution_function}),
\begin{equation}
\dot{m}_i = \dot{M}[F(z_{i+1}) - F(z_i)]
\label{eq:settling_rate}
\end{equation}
and set $\dot{m}_n = 0$. We model settling as a process that extracts mass from the convection zone and deposits it over a range of depths without any mixing in between (see Sec.~\ref{sec:model_physics}). Therefore the whole lithium content of $\dot{m}_i$ gets into the $i$-th sub-layer, and the lithium settling rate there is
\begin{equation}
R_{\trm{s},i} = A_n\frac{X\dot{m}_i}{m_\trm{p}},
\label{eq:rsi}
\end{equation}
where the hydrogen mass fraction $X$ and proton mass $m_\trm{p}$ are used to obtain the settling rate of hydrogen atoms, $X\dot{m}_i/m_\trm{p}$. The lithium abundance in the convection zone, $A_n$, then converts the hydrogen settling rate to the lithium settling rate. The advection part of Eq.~\ref{eq:burning_equations} refers to the transport of lithium by the upflow, which causes a mass flux of \mbox{$\sigma_i = \dot{M}F(z_i)$} through the bottom of the $i$-th sub-layer. This corresponds to a flux of lithium atoms of $A_{i-1}X\sigma_i/m_\trm{p}$ (coming from the $(i-1)$-st sub-layer). The rate of advective transport $R_{a,i}$ is then the difference between the inflow into and the outflow from the $i$-th sub-layer,
\begin{equation}
R_{\trm{a},i} = A_{i-1}\frac{X\sigma_i}{m_\trm{p}} - A_i\frac{X\sigma_{i+1}}{m_\trm{p}},
\label{eq:rai}
\end{equation}
which is negative.

Equation~\ref{eq:burning_equations} can be expressed in terms of abundances in the following way. First, insert Eqs.~\ref{eq:rbi}, \ref{eq:rsi} and \ref{eq:rai} into Eq.~\ref{eq:burning_equations} and divide each resulting equation by the corresponding hydrogen number density $N_{\trm{H}, i}$. Second, notice that $\sigma_i = \sum_{k=1}^{i-1} \dot{m}_k$ and define a discrete version of $\tau_\trm{r}$ as $\tau_{\trm{r}, i} = M_i/\dot{m}_i$, where the mass of the $i$-th sub-layer (per unit area) is $M_i = N_{\trm{H}, i}\,m_\trm{p}/X$. Third, rearrange terms to obtain
\begin{equation}
\frac{\diff \ln A_i}{\diff t} = -\frac{1}{\tau_{\trm{b}, i}} + \frac{A_n/A_i - 1}{\tau_{\trm{r}, i}} - \left(1 - \frac{A_{i-1}}{A_i}\right) \sum_{k = 1}^{i - 1}\frac{M_k}{M_i}\frac{1}{\tau_{\trm{r},k}}.
\label{eq:burning_equations_2}
\end{equation}
The physical effect of each of the three terms on the right-hand side of Eq.~\ref{eq:burning_equations_2} can now be seen easily. The first one exponentially destroys lithium on the local burning time scale $\tau_{\trm{b},i}$. The second term strives to equalise the local lithium abundance $A_i$ with that of the convection zone. This, as a direct effect of settling, happens on the local recycling time scale $\tau_{\trm{r},i}$. The last term describes how $A_i$ tends to approach $A_{i-1}$, i.e. the slow rising of the lithium stratification due to the upflow induced by settling. Its strength depends on the total `speed' ($\sim 1/\tau_\trm{r}$) of settling beneath the $i$-th sub-layer. 

% **********************************************************************
% *                             Results                                *
% **********************************************************************
\section{Results}
\label{sec:results}

Numerical simulations of the solar photosphere show that the entropy contrast between the upflow and a typical downflow is $\delta s = 1.8\times 10^8$\,erg\,K$^{-1}$\,g$^{-1}$ \citep[see Fig.~29 in][]{SteinNordlund98}. We set the maximal entropy contrast in the mass flux distribution to this value, which corresponds to $(\delta \hat{s})_\trm{max} = 1.3$ in our dimensionless units. We use \mbox{$(\delta \hat{s})_\trm{min} = 1.0\times 10^{-6}$} for the minimal entropy contrast in the mass flux distribution. We put the reference point to \mbox{$r = 0.50 R_\odot$} in the standard solar model (SSM), which is slightly deeper than the bottom of the settling layer for our choice of $(\delta \hat{s})_\trm{max}$. The gravitational acceleration is set such that the mass of the region with $T > 2.5\times 10^6$\,K in our model (with $\dot{M}=0$) is close to the corresponding value from the SSM.\footnote{ This of some importance for the lithium-burning calculation. Equation~\ref{eq:tau_r} shows that $\tau_\trm{r} \propto 1/g$. } We also adjust the opacity parameter $\alpha$ such that the entropy difference between the Schwarzschild boundary and the reference point matches the value from the SSM (again with $\dot{M} = 0$). 
\begin{figure}
\centering
\includegraphics[width=9cm]{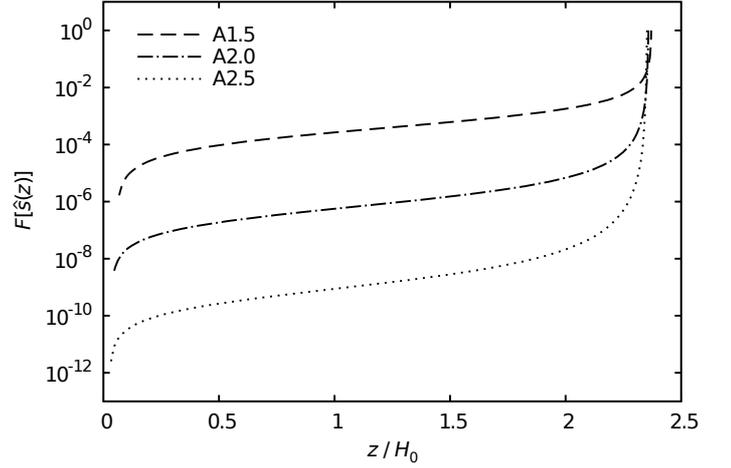}
\caption{The cumulative mass flux distributions in the A-type settling models plotted as functions of height. $F\left[\hat{s}(z)\right]$ is the fraction of the mass flux $\dot{M}$ that settles below the height level $z$. The height $z$ is counted upwards from a point beneath the settling layer. }
\label{fig:distributions_A}
\end{figure}
\begin{table}
\centering
\resizebox{\columnwidth}{!}{%
\begin{tabular}{c c c c c c}
\hline\hline
Model id.\rule[-0.75ex]{0pt}{3.25ex} & $\hat{\mathcal{F}}_\trm{conv,\,sb}$ & $\beta$ & $\dot{M}$ [g\,cm$^{-2}$\,s$^{-1}$] & $\max\left(\delta c_\trm{s}^2/c_\trm{s}^2\right)$ & \multicolumn{1}{c}{$\log_{10}\left[A_\trm{cz}(t_\odot)\right]$} \\
\hline
A1.5\rule{0pt}{2.5ex} & $10^{-2}$ & $1.5$ & $4.5\times 10^{-3}$ & $8.7\times 10^{-3}$ & \mbox{$<-10$} \\
A2.0 & $10^{-2}$ & $2.0$ & $4.4\times 10^{-1}$ & $2.2\times 10^{-3}$ & $-4.8$ \\
A2.5 & $10^{-2}$ & $2.5$ & $2.2\times 10^{0\phantom{-}}$ & $4.0\times 10^{-5}$ & $-0.091$ \\
B1.5 & $10^{-1}$ & $1.5$ & $4.9\times 10^{-2}$ & $8.7\times 10^{-2}$ & \mbox{$<-10$} \\
B2.0 & $10^{-1}$ & $2.0$ & $4.5\times 10^{0\phantom{-}}$ & $2.2\times 10^{-2}$ & \mbox{$<-10$} \\
B2.5\rule[-0.75ex]{0pt}{2.5ex} & $10^{-1}$ & $2.5$ & $2.2\times 10^{1\phantom{-}}$ & $3.7\times 10^{-4}$ & $-0.72$ \\
\hline
\end{tabular}%
}
\caption{Properties of the six settling models.}
\label{tab:models}
\end{table}
 
We first show the typical behaviour of settling on a set of six models (see Table~\ref{tab:models}), roughly sampling the corner of the parameter space that is likely to be relevant. We use three different values of $\beta$ and, instead of setting $\dot{M}$ to any particular value, we adjust it iteratively in order to reach certain values of the convective flux at the Schwarzschild boundary, \mbox{$\hat{\mathcal{F}}_\trm{conv,\,sb} \equiv 1 - \hat{\mathcal{F}}_\trm{rad,\,sb}$} (see Table~\ref{tab:models}). This is motivated by the fact that our mass flux distributions (see Fig.~\ref{fig:distributions_A}) are to represent a low-entropy tail {\em appended} to a distribution of MLT-like flows, which do not appear in our model. Since the $F(\hat{s})$ distributions have a sharp peak close to the entropy values predicted by the MLT, one can expect the convective flux induced by such set of downflows to be a non-negligible (although not precisely known) fraction of the total flux. Therefore we use $\hat{\mathcal{F}}_\trm{conv,\,sb} = 0.01$ and $\hat{\mathcal{F}}_\trm{conv,\,sb} = 0.1$ in A-type and B-type models, respectively (see Table~\ref{tab:models}). The cumulative mass flux distributions (Fig.~\ref{fig:distributions_A}) show that in all cases only a tiny fraction of the input mass flux reaches substantial depths. We do not show the distributions of the B-type models, because their profiles are almost the same as the ones shown, but shifted along the $z$ axis.
\begin{figure}
\centering
\includegraphics[width=9cm]{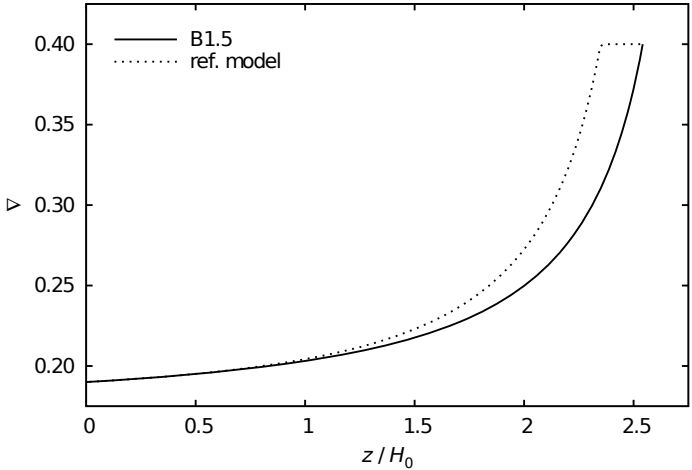}
\caption{The temperature gradient in the model B1.5 (solid line) compared with the reference model (dotted line). In the convectively unstable part of the reference model $\nabla = \nabla_\trm{ad} = 0.4$ (see Sec.~\ref{sec:results}).}
\label{fig:nabla_B1.5}
\end{figure}

We compare the properties of our settling models with a reference one having $\dot{M} = 0$ (i.e. without settling). The comparison is made at the same geometrical height $z$, normalised to the pressure scale height at the reference point, $H_0 = \mathbb{R}T_0/g$. The convective part of the reference model extends a bit deeper than in the settling models; i.e., the settling process {\em reduces} the depth of the convection zone (cf. discussion in Sec.~\ref{sec:discussion}). We set $\nabla = \nabla_\trm{ad} = 0.4$ in the convectively unstable part of the reference model to make the comparison of thermodynamic quantities possible. 

Figure~\ref{fig:nabla_B1.5} compares the temperature gradient in the model B1.5 (the strongest settling) to the reference one. It clearly shows that the temperature gradient in the settling models has to decrease in order to reduce the radiative flux, as expected in Sec.~\ref{sec:model_physics}. We can see that models with settling indeed reach convective instability higher up in the stratification. It is also evident from Fig.~\ref{fig:nabla_B1.5} that settling preserves the discontinuity in the slope of $\nabla$ (hence in the second derivative of the sound speed) at the Schwarzschild boundary, which has consequences for helioseismology.
\begin{figure*}
\centering
\includegraphics[width=17cm]{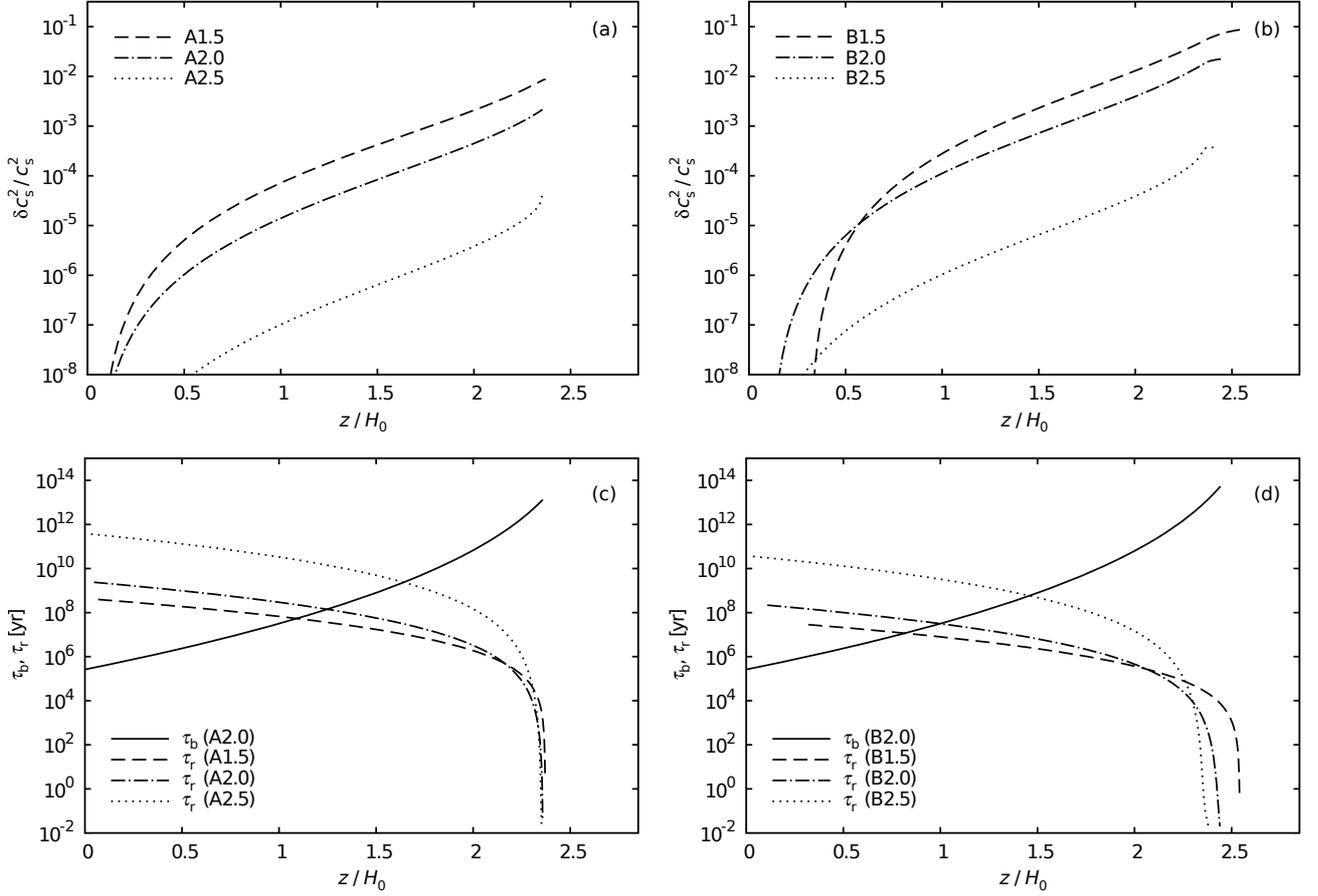}
\caption{(a,\,b) The differences in the squared sound speed between the settling models and the reference model. The right end of each curve marks the position of the Schwarzschild boundary. (c,\,d) The lithium-burning time scale $\tau_\trm{b}$ and the recycling time scale $\tau_\trm{r}$ in the models. $\tau_\trm{b}$ is only plotted for the models with $\beta = 2.0$, because it is very similar in the other ones.}
\label{fig:soundspeed_timescales}
\end{figure*}

Figures~\ref{fig:soundspeed_timescales}a,b show the relative {\em increase} in the squared sound speed due to settling,
\begin{equation}
\frac{\delta c_\trm{s}^2}{c_\trm{s}^2} = \frac{c_\trm{s}^2 - c_\trm{s,\,r}^2}{c_\trm{s,\,r}^2},
\end{equation}
where $c_\trm{s}$ is the sound speed in the settling model and $c_\trm{s,\,r}$ is the sound speed in the reference one. The maximum values of $\delta c_\trm{s}^2/c_\trm{s}^2$ reached by our models are also listed in Table~\ref{tab:models}.

Plots of the recycling time scale in Figs.~\ref{fig:soundspeed_timescales}c,d show that a considerable fraction of the settling layer gets mixed with the convection zone on the time scale of the Sun's lifetime. Such mixing could alter the chemical gradients caused by slow element diffusion and change the sound speed profile with respect to the SSM. 

The conditions for lithium burning can be qualitatively judged using Figs.~\ref{fig:soundspeed_timescales}c,d, which show the recycling time scale $\tau_\trm{r}$ (Eq.~\ref{eq:tau_r}) and the lithium-burning time scale $\tau_\trm{b}$ (Eq.~\ref{eq:tau_b}) as functions of height. Since they both change by many orders of magnitude, we estimate the overall extent of lithium depletion in the convection zone by integrating Eq.~\ref{eq:burning_equations_2} with the initial condition $A_i = 1$,\ $i = 1,\,2,\,\dots,\,n$, so that our calculations show the {\em relative} change in the lithium abundance with respect to the initial one. The integration is stopped at $t = t_\odot = 4.5\times 10^9$\,yr. The strongly varying extent of lithium depletion in the convection zone $A_\trm{cz} \equiv A_n$, listed in Table~\ref{tab:models}, shows the extreme sensitivity of the lithium depletion rate to $\tau_\trm{r}$, $\tau_\trm{b}$ and thus to $\beta$. We only present models for one value of $(\delta \hat{s})_\trm{max}$, but the consequences of changing this parameter can be judged using Figs.~\ref{fig:soundspeed_timescales}c,d.

First, consider a model with $\tau_\trm{r} \gg t_\odot$ at the bottom of the settling layer. Under this condition, the convection zone does not ,,feel'' the bottom of the settling layer, because the material exchange between them over the Sun's lifetime is negligible. The lithium depletion cannot depend much on $(\delta \hat{s})_\trm{max}$ in this case. If, on the other hand, we take a model with $\tau_\trm{r} \ll t_\odot$ at the bottom of the settling layer, the exchange of mass with the convection zone is efficient. The rate of lithium depletion in the convection zone is sensitive to the maximal depletion rate in the settling layer, i.e. the value at its bottom. Such a model must therefore be sensitive to $(\delta \hat{s})_\trm{max}$ since this parameter determines the depth of the settling layer.
\begin{figure}
\centering
\includegraphics[width=9cm]{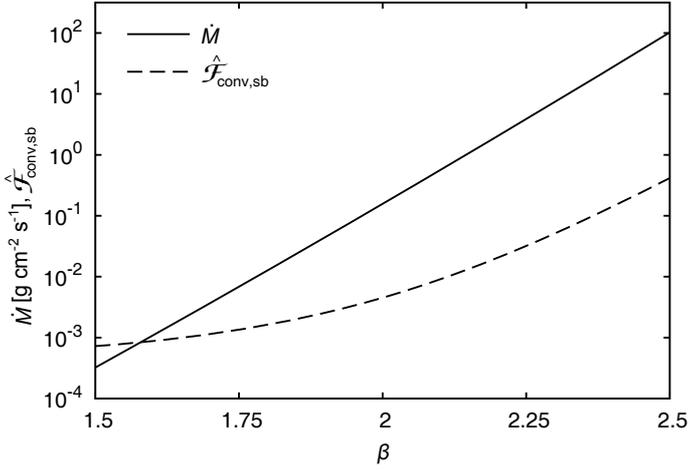}
\caption{The settling mass flux $\dot{M}$ and the convective flux at the Schwarzschild boundary $\hat{\mathcal{F}}_\trm{conv,\,sb}$ plotted as functions of $\beta$ in a set of models calibrated to produce the observed lithium depletion in the Sun.}
\label{fig:mdot_fconv_vs_beta}
\end{figure}

Observations tell us that the relative lithium depletion in the Sun is $\log_{10}\left[A_\trm{cz}(t_\odot)\right] = -2.21\pm 0.11$, see e.g. \citet{AsplundEtal09}. We have calibrated a set of models (not shown in Table~\ref{tab:models}) by adjusting $\dot{M}$ at any given value of $\beta$ until they all predicted the observed lithium depletion. These models use the same values of $(\delta \hat{s})_\trm{min}$ and $(\delta \hat{s})_\trm{max}$ as the ones listed in Table~\ref{tab:models}. The derived values of $\dot{M}$ and $\hat{\mathcal{F}}_\trm{conv,\,sb}$ are plotted in Fig.~\ref{fig:mdot_fconv_vs_beta} as functions of $\beta$. The recycling time scale at the bottom of the settling layer in this set of models is $\sim 6.9\times 10^9$~yr with only $16\%$ variation over the range of $\beta$ plotted in Fig.~\ref{fig:mdot_fconv_vs_beta}. This is more than the age of the Sun, so that the burning-calibrated models are only mildly sensitive to the assumed value of $(\delta \hat{s})_\trm{max}$. The peak values of $\delta c_\trm{s}^2/c_\trm{s}^2$ in these models range from $6\times 10^{-4}$ at $\beta = 1.5$ to $2\times 10^{-3}$ at $\beta = 2.5$. 

% **********************************************************************
% *                      Summary and discussion                        *
% **********************************************************************
\section{Summary and discussion}
\label{sec:discussion}

We have shown that there are several processes contributing to overshooting under a convective stellar envelope, which cover a wide range of depth and time scales. The `ballistic' form of overshooting  acts on the dynamical time scale in a shallow boundary layer. The process of convective penetration needs more time to spread but reaches deeper. Convective settling, potentially the deepest-reaching overshooting process, only plays a role on the time scale of the star's lifetime. Finally, gyroscopic pumping and internal gravity waves may also contribute to slow mixing.

We have isolated the process of convective settling, in which a small fraction of the cold photospheric material is assumed to maintain some of its low entropy until it has settled at its neutral-buoyancy level in the stable stratification beneath the convection zone. The typical entropy contrast observed in photospheres of solar-type stars is strong enough so that settling can in priniciple penetrate the lithium-burning layers (Spruit 1997). However, an order-of-magnitude estimate shows that the mass flux of this material required to explain the observed lithium depletion in the Sun on the nuclear time scale is tiny (a fraction of order $10^{-7}$ of the mass flux in the downflows at the surface). Effects as weak as this cannot be measured from direct hydrodynamic simulations.

On its way down, the cold material mixes by entrainment with its surroundings, a process that cannot be captured realistically with existing means. The likely outcome at the base of the convection zone is a distribution of entropy contrast in the downflows, with most of the mass flux near the value corresponding to a mixing-length estimate, but with a tail of unknown shape extending to much lower entropy values. In the absence of further information, we have parametrised this tail by a power law. 

We use a one-dimensional model that treats the settling material as a depth-dependent source of mass in the layers below the base of the convection zone. The response of this region is a slow upward flow. Accompanying it, there is a thermal adjustment by radiative diffusion. The result is a slow, depth-dependent circulation of material. The free parameters of the entropy distribution model are adjusted such that lithium depletion takes place on the observed time scale.

The results show that the radiative flux and temperature gradient decrease to compensate for a positive (upward) convective flux caused by settling. This decrease in $\nabla$ leads to a slight reduction of the convection zone's depth. The settling process is calculated only as a perturbation of a precalculated solar model, however; i.e., we do not model the evolution of the Sun with settling included. In a self-consistent and properly calibrated stellar-evolution model the change in the depth of the convection zone might actually have the opposite sign.

The resulting change in the sound speed profile due to settling is rather small and is concentrated in a thin layer below the Schwarzschild boundary. The calculation, however, only includes the direct effect of the settling process. It does not include the changes in the Sun's structure during its evolution on the main sequence. It is possible that this will redistribute the structural changes due to settling over a larger portion of its radius.

Another factor not included in our calculations is the stratification of helium concentration below the convection zone. It is caused by gravitational settling (also called `diffusion') and known to significantly influence the solar sound-speed profile \citep[see][and references therein]{DalsgaardMauro07}. This helium-concentration gradient could be modified by the mixing induced by convective settling.

As Fig.~\ref{fig:nabla_B1.5} shows, settling preserves the discontinuity in the second derivative of the sound speed. That the helioseismic observations favour smoother sound-speed profiles \citep{DalsgaardEtal11} is evidence for the existence of an additional overshooting mechanism acting closer to the boundary of convective instability than the settling process studied here. 

The predicted lithium depletion changes by many orders of magnitude, as a function of the two model parameters (see Table~\ref{tab:models}). This is a natural consequence of the high temperature sensitivity of the burning reaction, combined with rapid changes of the mass flux in the lithium-burning layers as we change the slope $\beta$ of the mass flux distribution. Therefore we use the observed lithium depletion in the Sun to constrain our model. This constraint yields a dependence of the total input mass flux $\dot{M}$ on slope $\beta$, reducing the number of free parameters to one (see Fig.~\ref{fig:mdot_fconv_vs_beta}). The linearity of $\log_{10}\dot{M}(\beta)$ comes as no surprise if we inspect Fig.~\ref{fig:distributions_A} in detail. We see that the $F\left[\hat{s}(z)\right]$ distributions at significant depths are self-similar and they apparently shift in proportion to $\beta$ (in the logarithmic space of Fig.~\ref{fig:distributions_A}); i.e., a change in $\beta$ can be directly translated to an equivalent change in $\dot{M}$ if the overall extent of lithium depletion is fixed. Figure~\ref{fig:mdot_fconv_vs_beta} also shows that the most relevant values of $\beta$ lie somewhere in the interval $(1.5,\, 2.5)$ or even $(2.0,\, 2.5)$. At higher values we could not meet the lithium depletion constraint because such models would require negative radiative flux at the top of the settling layer. At the low end, the convective flux becomes a negligible fraction of the total flux, and it gets difficult to interpret our power law as a tail of some more general mass flux distribution. 

One might ask how our strongly simplified model compares to other models of overshooting. The most striking difference is that usually some form of the MLT is used to provide estimates of the velocities and entropy fluctuations at the boundary of the convection zone \citep[e.g.][]{Roxburgh65, SaslawSchwarzschild65, ShavivSalpeter73, vanBallegooijen82, PidatellaStix86, Zahn91}, whereas we explicitly add the hypothesised low-entropy flows from the photosphere. Perhaps the closest to our ideas are the works of \citet{Rempel04} and \citet{SchmittEtal84}, who model the non-local convection by plumes.

The approach of modelling higher-order correlations in a turbulent field \citep[e.g.][]{Kuhfuss86, XiongDeng01, MarikPetrovay02, DengXiong08} tends to produce much deeper overshooting zones than the models above (depending on the value of the free parameters of the models). This can be traced to the fact that these models lack an essential aspect of the transition 
between convection and the stable interior. While velocity {\em amplitudes} vary rather smoothly across the boundary, their mixing effect varies strongly. In the convective region, the flows are of the efficiently mixing, overturning kind. In the stable part, however, the flows take the form of internal waves, which have a very weak mixing effect. These models also cannot capture the process of convective settling since the rare low-entropy downflows are not present in this picture.

In summary, we have shown that the convective settling process studied here can in principle explain long-term lithium depletion in the Sun and solar-type stars. This can be tested further by applying the model, calibrated to the lithium depletion observed in the Sun, to stars of different masses and ages. This will be pursued in a follow-up paper. 

\section*{Acknowledgements}

We thank the anonymous referee for critical comments that improved the discussion section and the overall presentation of the text.

\end{document}